\documentclass[galaxies,review,accept,moreauthors,pdftex,10pt,a4paper]{mdpi} 

\def\gsim{\mathrel{\raise.5ex\hbox{$>$}\mkern-14mu
             \lower0.6ex\hbox{$\sim$}}}
\def\lsim{\mathrel{\raise.5ex\hbox{$<$}\mkern-14mu
             \lower0.6ex\hbox{$\sim$}}}

\firstpage{1} 
\pubvolume{xx}
\issuenum{1}
\articlenumber{5}
\pubyear{2019}
\copyrightyear{2019}
\history{Received: date; Accepted: date; Published: date}
\updates{yes} % If there is an update available, un-comment this line

\Title{Generation and Transport of Magnetic Flux in Accretion--Ejection Flows}

\Author{{Ioannis} Contopoulos $^{1,2,\dagger}$
}   %Please carefully check the accuracy of names and affiliations. Changes will not be possible after proofreading.

\AuthorNames{Ioannis Contopoulos}

\address{%
$^{1}$ \quad Research Center for Astronomy and Applied Mathematics, Academy of {Athens, Athens 11527, Greece}; icontop@academyofathens.gr; Tel.: +30-210-6597167
\\  %Please check the added information.
$^{2}$ \quad National Research Nuclear University (Moscow Engineering Physics {Intitute), 31 Kashirskoe Highway, 115409 Moscow, Russia}} %Please check the added information.

\firstnote{Current address: 4 Soranou Efessiou Str., Athens 11527, Greece.} 

\abstract{Astrophysical accretion flows are associated with energetic emission of radiation and outflows (winds and jets). Extensive observations of these two processes in X-ray binary outbursts are available. A convincing understanding of their dynamics remains, however, elusive. The main agent that controls the dynamics is believed to be a large scale magnetic field that threads the system. We propose that during the quiescent state, the field is held in place by a delicate balance between inward advection and outward diffusion through the accreting matter. We also propose that the source of the field is a growing toroidal electric current generated by the {aberrated} radiation pressure on the innermost plasma electrons in orbit around the central black hole. This is the astrophysical mechanism of the Cosmic Battery. When the return magnetic field outside the toroidal electric current diffuses through the surrounding disk, the disk magnetic field and its associated accretion rate gradually increase, thus leading the system to an outburst. After the central accretion flow approaches equipartition with radiation, it is disrupted, and the Cosmic Battery ceases to operate. The outward field diffusion is then reversed, magnetic flux reconnects with the flux accumulated around the central black hole and disappears. The magnetic field and the associated accretion rate slowly decrease, and the system is gradually driven back to quiescence. We conclude that the action (or inaction) of the Cosmic Battery may be the missing key that will allow us to understand the long-term evolution of astrophysical accretion--ejection flows.%NOTE: Please confirm this word.
}

\begin{document}
%%%%%%%%%%%%%%%%%%%%%%%%%%%%%%%%%%%%%%%%%%
%% Only for the journal Gels: Please place the Experimental Section after the Conclusions

%%%%%%%%%%%%%%%%%%%%%%%%%%%%%%%%%%%%%%%%%%
%\setcounter{section}{-1} %% Remove this when starting to work on the template.
\section{Accretion--Ejection Flows around Astrophysical Black Holes}

Active galactic nuclei (AGN), X-ray binaries (XRB) and other energetic astrophysical sources are believed to be powered by the infall (accretion) of gaseous matter (plasma) into a central black hole. The infall proceeds as a  rotating disk along which matter gradually releases enormous amounts of gravitational energy in the form of energetic outflows (winds and jets) and radiation across the full electromagnetic spectrum.

In their seminal paper \citep{SS73} (hereafter SS73), Shakura     and     Sunyaev proposed that the structure and radiation spectrum of the accretion disk depend mainly on the matter accretion rate $\dot{M}_{\rm disk}$ in the disk. They went on to calculate the radiation spectrum as a superposition of black-body spectra emitted as matter locally converts gravitational energy into radiation at all radii in the disk. Their model serves as the standard model of accretion flows to the present day. It is interesting that SS73 realized that high-energy radiation can evaporate the gas and counteract the matter inflow in the disk. They concluded that, at high accretion rates, most of the gas will outflow along the way from the outer to the central regions of the disk, and only a small fraction will accrete into the central black hole. The~observational fact that accretion disks indeed generate outflowing winds and jets from their surfaces nevertheless remains rather surprising.

Morover, it has been observed that, in many cases, the kinetic power of the outflow exceeds the total luminosity of the disk by several orders of magnitude \citep{Getal14, Setal10, Aetal10}. This may be explained by the presence of an external agent that removes angular momentum and thus also gravitational energy from the accretion flow. In the SS73 picture, angular momentum is removed through viscous stresses in the disk, and, therefore, gravitational energy is released locally at all radii as high-energy radiation. Recent observations suggest that angular momentum is removed via a large scale magnetic field that threads the accretion and ejection flows. In doing so, gravitational energy is transformed into the kinetic energy of the outflow, thus there is no need for it to be radiated locally in the disk. This action of the large scale magnetic field suppresses the total radiation from the disk and modifies its high-energy spectrum \citep{Fetal06, Metal18a, Metal18b}.

The study of accretion flows is very complicated by itself. Now that we have concluded that accretion flows co-exist and as it seems do depend strongly on ejection flows, one needs to study accretion and ejection as one coherent process. Over the past couple of decades, several works contributed to the development of our understanding of accretion{--}ejection flows (e.g., \citep{FP93, FP95, CK02, Zetal07, MFZ10, Setal12, SF16, Qetal17, QFV18}). This may not be the common view of the community, but we consider the work of Ferreira and collaborators as the most comprehensive. Due to the high degree of complexity of the problem, their treatment of the accretion flow has been similar to that of SS73. Accretion proceeds in the form of a more-or-less standard Shakura--Sunyaev-type disk, and ejection takes place as a gradual small perturbation above its surface. Let us denote with $\dot{M}_{\rm disk}(r)$ the accretion rate at radius $r$ in the disk, and $\dot{M}_{\rm wind}(r)$ the total outflow rate enclosed within radius $r$ in the wind. We   assume that the accretion--ejection structure extends from an internal boundary near the black hole horizon at radius $r_{\rm in}$ to an outer boundary at radius $r_{\rm out}$. We   also assume that the densities $\rho_{\rm disk}(r)$ in the disk and $\rho_{\rm wind}(r)$ at the base of the outflow vary as power laws with distance, namely, %NOTE: Please confirm - vs. -- vs. --- vs. $-$ throughout.
\begin{equation}
\rho_{\rm wind}(r)\propto \rho_{\rm disk}(r)\propto r^{-p}
\end{equation}
(in a radially self-similar configuration all densities have the same power-law dependence), and that both the ejection velocity $v_{z}$ at the base of the wind and the accretion velocity $v_{r}$ in the disk follow Keplerian profiles
\begin{equation}
v_{z}(r)\propto v_{r}(r)\propto r^{-1/2}
\end{equation}

Finally, we   assume that the disk scale-height $h\equiv r c_s/v_{\rm K}$ is proportional to $r$ ($c_s$ is the speed of sound at the midplane of the disk, $v_{\rm K}\equiv (GM/r)^{1/2}$ is the Keplerian velocity, and $M$ is the mass of the central black hole). Putting all the above together, we obtain that
\begin{eqnarray}
&& \dot{M}_{\rm disk}=2\pi r h v_{r}\rho_{\rm disk}\propto r^{1.5-p}\\
&& \dot{M}_{\rm wind}=\int_{r_{\rm in}}^{r_{\rm out}}
2\pi r v_{z}\rho_{\rm wind}{\rm d}r
\end{eqnarray}
and since mass conservation requires that ${\rm d}\dot{M}_{\rm wind}/{\rm d}r={\rm d}\dot{M}_{\rm disk}/{\rm d}r$, we obtain
\begin{equation}
\xi\equiv 1.5-p=\frac{\rho_{\rm wind}}{\rho_{\rm disk}}  \frac{r}{h}\frac{v_{z}}{v_{r}}
\end{equation}
where  $\xi$ is the local ejection efficiency parameter of \citep{FP93}. In general, we expect that $v_{z}$ at the base of the wind is comparable to $v_{r}$ in the disk. When $\dot{M}_{\rm disk}\approx$~const., $\xi\approx 0$, $\rho_{\rm wind}\ll \rho_{\rm disk}$, and, therefore, the flow is dominated by a strong accretion disk, a small fraction of which outflows to infinity in the form of a wind. On the contrary, when $\xi$ differs significantly from zero, $\rho_{\rm wind}$ may be a significant fraction of $\rho_{\rm disk}$, and, therefore, $\dot{M}_{\rm wind}(r_{\rm out})\approx \dot{M}_{\rm disk}(r_{\rm in})(r_{\rm out}/r_{\rm in})^\xi\gg \dot{M}_{\rm disk}(r_{\rm in})$. Ferreira and collaborators concluded that the ejection efficiency parameter must be very small \citep{FP93, FP95}. {More specifically}, they concluded that, in cold flows, the disk winds are only a small perturbation of the accretion process. For $\xi\approx 0$, the wind above the disk may be described by the Blandford     and     Payne~(hereafter BP82) radially self-similar wind solution with $B\propto r^{-5/4}$ and $\rho\propto r^{-3/2}$ \citep{BP82}. This has been the canonical model of disk winds for almost three decades now.

It is very interesting that BP82-type solutions do not extend to infinity, but are diverted toward the axis beyond some distance \citep{C92, CL94}. Several years later, the scaling of the density in the wind with radial distance was generalized to $\rho_{\rm wind} \propto r^{-p}$ \citep{CL94, KK94}. Contopoulos showed that, among radially self-similar {solutions}, only the ones with $p\leq 1$ extend to infinity in the form of cylindrically collimated jets. What is most interesting though is that the analysis of observations of X-ray absorption lines in extended winds from AGN and XRB by Fukumura and collaborators \citep{Fetal10a, Fetal10b, Fetal14, Fetal15, Fetal17} have shown that such extended disk winds may be modeled best by power-law exponents $p$ closer to 1 than 1.5 (e.g., $p\approx 1.2$ in the wind of GRO J1655-40 \citep{Fetal17}). This observational result prompted {\citep{Chaketal16}} to propose that disk winds may actually be generated by a ``warm'' (instead of ``cold'') disk. On the other hand, {\citep{CKF17}} even considered the possibility that the magnetic field in the disk is not held in place by a balance between inward advection of the field by the accreting matter and outward diffusion of the field through the disk. Our present understanding is that a value of $p$ different from $1.5$ does not necessarily imply flux imbalance. We expect that a density scaling with $p\neq 1.5$ may be compatible with $\dot{M}_{\rm disk}\approx \mbox{const.}$ if $h\propto r^{p-1/2}$ (and not $\propto r$). In the present work, we assume that $p=1.5$, but our conclusions should remain valid even for a different density scaling in the wind (although this has not been formally verified).

The issue of the transport of magnetic flux through the disk is not new. The assumed presence of the large-scale magnetic field threading the disk required the efficient inward transport of magnetic flux. This issue has been strongly debated over the years (e.g., \citep{vB89, LPP94, LRN94}). What is even more perplexing is that  the most recent state-of-the-art numerical simulations of magnetized accretion--ejection flows performed by the Harvard group (e.g., \citep{TNMcK11, McKTB12})  reach a so-called {Magnetically Arrested Disk (MAD)} state where, seemingly, accretion proceeds continuously, whereas the total magnetic flux accumulated over the central black hole saturates to a limiting maximum value. However, if one studies these numerical simulations more closely, one realizes that every parcel of accreting matter brings the magnetic flux associated with it very close to the black hole horizon. Then, right before that parcel of matter plunges into the horizon, it gets rid through reconnection of the magnetic flux that it carried all along. Matter continues to accrete, but magnetic flux accumulates in the vicinity of the black hole, continuously reducing the average density of the accreting matter. It is straightforward to see that, over time in these simulations, $\dot{M}_{\rm disk}(r_{\rm in})$ and the total magnetic flux $\Psi_{\rm BH}$ accumulated onto the central black hole indeed both remain unchanged on average. However, the density of the surrounding disk continuously decreases with time as one can see by the color becoming more and more yellow in the top panel of Figure~1 in \citep{TNMcK11}. The duration of these simulations may be very long in numerical terms (several tens of thousands of dynamical times $GM/c^3$), but in actual physical terms, the time scale is tiny. To our understanding, the issue of magnetic flux transport in accretion {disks} remains open, and the numerical simulations of the Harvard group did not produce a convincing answer.%NOTE: Please define abbreviations where appropriate.

The origin of the magnetic field itself remains an open question.  {\citep{McKTB12}} argued that ``observations do show patches of coherent magnetic flux surrounding astrophysical systems that can feed black holes''. They reached the conclusion that even if only about 10 percent of it manages to accrete via the accretion disk, it would be enough to generate all the interesting accretion--ejection phenomena that we are investigating (see references in their paper). Here, we   discuss another possibility, namely that constant-polarity flux is generated by the Poynting--Robertson drag effect on the plasma electrons in a so-called Cosmic Battery around the central black hole \citep{CK98, C15}. As we show below, the astrophysical implications of this effect, first proposed by Contopoulos     and     Kazanas more than twenty years ago, may be the missing key needed to understand the long time-evolution of accretion--ejection structures.

\section{The Origin of the Magnetic Field: A Cosmic Battery}

Let us discuss first how radiation acts on a plasma. In non-relativistic dynamics, radiation is introduced as an extra term in the equation of motion of the plasma, namely
\begin{equation}\label{eqmotionplasma}
\rho\frac{{\rm d}{\bf v}}{{\rm d}t}=\dots + \frac{\rho}{m_p}{\bf f}_{\rm rad}\ ,
\end{equation}
where  $\rho$ and ${\bf v}$ are the plasma matter density and velocity, respectively, and $m_p$ is the mass of the proton (for ease of presentation, we   assume a simple electron--proton plasma).  It is well known that radiation acts on the plasma electrons, and {\it much less} on the plasma protons, and in fact ${\bf f}_{\rm rad}$ is the radiation force per electron. Thus, how is it possible that radiation contributes to the dynamics of the plasma as a whole, e.g., how is it possible that radiation holds stars from collapsing under their own weight? The answer is that, in the presence of radiation, {\it an inductive electric field} ${\bf E}$ {\it develops in the interior of the plasma}. Why is this electric field important becomes clear when we consider the equations of motion not only of the protons, but also of the electrons. In the presence of radiation, the equation of motion of the protons contains an extra term
\begin{equation}\label{eqmotionprotons}
m_p\frac{{\rm d}{\bf v}_p}{{\rm d}t}=\dots + e{\bf E}\ ,
\end{equation}
and the equation of motion of the electrons also contains the radiation force term
\begin{equation}\label{eqmotionelectrons}
m_e\frac{{\rm d}{\bf v}_e}{{\rm d}t}=\dots + {\bf f}_{\rm rad} - e{\bf E}\ .
\end{equation}
{Here,} ${\bf v}_p, {\bf v}_e$ are the velocities of the protons and the electrons, respectively, $m_e$ is the mass of the electron, and $e$ is the magnitude of the electron charge. The velocities of the electrons and the protons do not differ much and  $m_e\ll m_p$, therefore, Equation~(\ref{eqmotionelectrons}) is equal to zero to a very good approximation.~Thus,
\begin{equation}\label{E}
{\bf E}=\frac{{\bf f}_{\rm rad}}{e}\ .
\end{equation}

This is how the radiation force appears in the equation of motion for the protons and the plasma as a whole. This effect is often ignored by the younger generation of researchers. Another way to explain this result is that, without the electric field, the radiation force would have disturbed the motion of the electrons so dramatically that an enormous electric current and an associated magnetic field would have appeared in the interior of the highly conducting astrophysical plasma. This is, however, not the case. It is not possible to ``simply turn on'' an electric current inside a highly conducting plasma. Maxwell's equations and in particular Faraday's induction law teach us that the plasma will react, and an induction electric field will appear in the direction opposite to the direction of the growing electric current that will counteract its growth. It is thus not correct to study the growth of the electric current and the associated magnetic field by considering the velocity difference between the electrons and the protons, and their number densities in the plasma. This approach ignores the inductive reaction of the plasma, and leads to wrong conclusions. 

The radiation force per electron is calculated in the rest frame of the electron. Even if the radiation field is isotropic, the moving electron absorbs photons coming from the direction opposite to its direction of motion, and then re-emits them isotropically, losing momentum in the process. This is how the electrons feel the radiation force. In relativistic tensor notation, the spatial components of the radiation force are equal to
\begin{equation}
f^i_{\rm rad}=\frac{\sigma_T F^i_{\rm rad}}{c}
\end{equation}
where  $\sigma_T$ is the electron Thomson cross-section  and $F^i_{\rm rad}$ is the radiation flux (flow of radiation energy per unit surface) as seen in the frame of the electron. The radiation flux components are given by the projection of the stress-energy tensor of the radiation $T^{\mu \nu}_{\rm rad}$ in the frame of the electron as
\begin{equation}
F^i_{\rm rad}=h^i_\nu T_{\rm rad}^{\mu \nu}u_{\mu}\ ,
\end{equation}
where, $u^\mu$ is the electron {4}-velocity which almost coincides with that of the plasma, and $h^\mu_\nu\equiv -\delta^\mu_\nu -u^\mu u_\nu$ is a tensor that projects opposite to the target electron {4}-velocity \citep{ML96}.%NOTE: Can this be changed to ``4''? Please confirm throughout.

\begin{figure}%[H]
\centering
\includegraphics[width=16 cm]{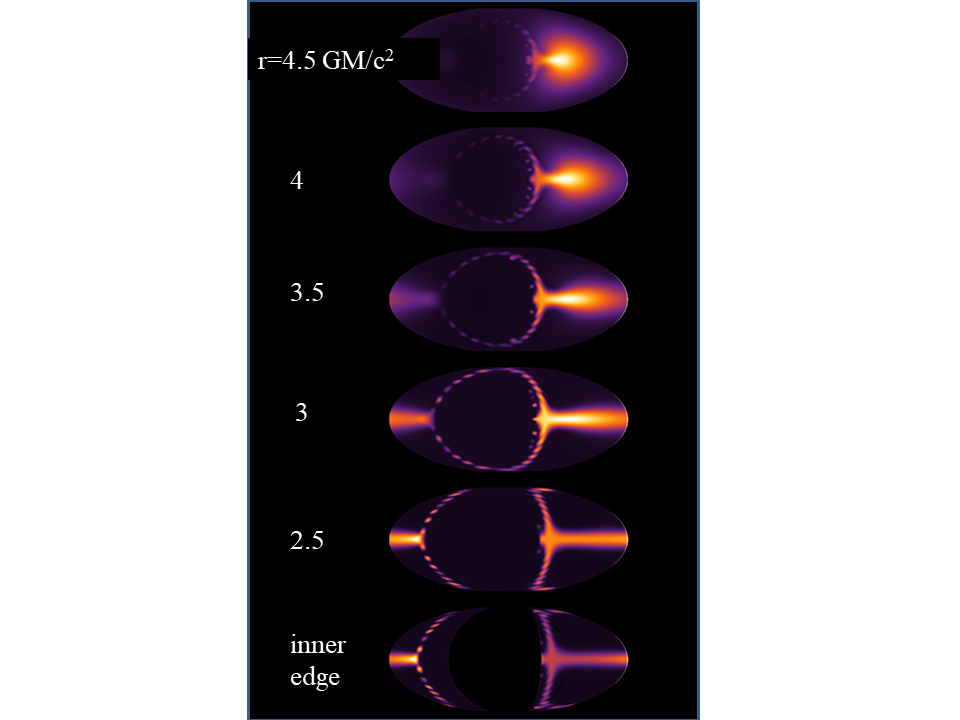}
\caption{Sky maps in Mollweide projection obtained by ray-tracing of the radiation field emitted by an equatorial torus as seen from an equatorial Keplerian observer at various distances $r$ in units {of} $GM/c^2$ (see {Footnote~1} for details). The central black hole and its associated ``light ring'' can {be discerned}. Black hole spin parameter $a=0.9$ M. The direction of rotation is to the right of the black hole. The color scale corresponds to the intensity of radiation.}
\label{FigureSkyMaps}
\end{figure}

The calculation of the radiation field in the vicinity of an astrophysical source of X-rays is complex and involves radiation transfer with absorption, emission, and detailed ray tracing. Another simpler approach is to consider the radiation field as a fluid, but this is not as accurate (see \citep{Cetal18} for details).  {\citep{KC14}} performed ray-tracing calculations over several optically thick equatorial distributions of matter (thin and thick disks, thick torus) extending beyond the Innermost Stable Circular Orbit (ISCO) around the central spinning black hole. Koutsantoniou (unpublished work)   recently extended this calculation over an optically thin radiation-emitting {torus.}\footnote{The distribution of matter in that torus is ad hoc, not a result of a numerical simulation. The torus has a circular cross section between $r_{\rm ISCO}$ and $2r_{\rm ISCO}$, and a density distribution that drops exponentially such that the optical depth from the center to the surface is taken to be equal to 2. The motion is everywhere Keplerian.} In Figure~\ref{FigureSkyMaps}, we show in Mollweide projection\footnote{The Mollweide or homalographic projection is a common map projection generally used for global maps of the world or the sky. The equator is represented as a straight horizontal line perpendicular to the central meridian in the direction of the {black hole}.} the view of the torus as seen from the frame of the plasma rotating with Keplerian velocities at various distances in the equatorial plane. The central black hole and its associated ``light ring'' can {be discerned}. The direction of rotation is to the right of the black hole. The region to the left of the black hole is the region opposite to the direction of the rotating plasma. It is interesting to notice here that inside about $3GM/c^2$, the radiation field becomes stronger along the direction of motion. This effect is due to the rotation of spacetime which forces most photons near the black hole horizon to reach the rotating plasma from behind. The normalized azimuthal radiation force per electron at the ISCO around a $5M_\odot$ Kerr black hole with various spin parameters $a$ are listed in the third column of Table~\ref{table1}. %FORMATTING: Footnotes.

We can obtain a crude estimate of the radiation force per electron in the idealized case of a central point radiation source emitting with luminosity $L$, and an electron in circular motion around it at distance $r$. In that case, 
\begin{equation}\label{PRforce}
{\bf f}_{\rm rad}=\frac{L\sigma_T}{4\pi r^2 ce}\hat{r}-\frac{L\sigma_T}{4\pi r^2 ce}\left(\frac{v^\phi}{c}\right)\hat{\phi}\ .
\end{equation}

The latter is the same expression as the well known Poynting--Robertson azimuthal drag force on dust grains in orbit around the  sun, only in that case the Thomson cross-section is replaced by the geometric cross-section of the grains \citep{P03, R37}. Obviously, the radiation field around an accreting rotating black hole is far more complex, as suggested by the calculated sky maps shown in Figure~\ref{FigureSkyMaps}.

\unskip

\begin{table}%[H]
\caption{Azimuthal Radiation Force per electron at the ISCO and Cosmic Battery Timescales.} \label{table1}
\centering
% \tablesize{} %% You can specify the fontsize here, e.g.,  \tablesize{\footnotesize}. If commented out \small will be used.
\begin{tabular}{cccc}
\toprule
%\textbf{
\boldmath{$a/M$}	& \boldmath{$r_{\rm ISCO}$} & \boldmath{$f_{\rm rad}^\phi$} &
\boldmath{$\tau_{\rm CB}$}\\
 &  \boldmath{$/(GM/c^2)$}	& \boldmath{$/(GMm_p/r_{\rm ISCO}^2)$} & \textbf{in Hours} \\
\midrule
0		& 6			& $-$0.11 & 39 \\
0.1	& 5.7			& $-$0.12 & 31 \\
0.2	& 5.3		& $-$0.13 & 25 \\
0.3	& 5.0		& $-$0.15 & 19 \\
0.4	& 4.6		& $-$0.18 & 15 \\
0.5	& 4.2			& $-$0.21 & 11 \\
0.6	& 3.8			& $-$0.27 & 8 \\
0.7	& 3.4			& $-$0.38 & 5 \\
0.8	& 2.9			& $-$0.59 & 3 \\
0.9	& 2.3			& +1.19 & 2 \\
0.92	& 2.2			& +1.46 & 2 \\
0.94	& 2.0			& +1.87 & 1 \\
0.96	& 1.8			& +2.58 & 1 \\
0.98	& 1.6			& +4.17 & 1 \\
\bottomrule
\end{tabular}
\end{table}

We {will} now discuss the growth of the magnetic field generated by the intense radiation and rotational velocity fields in the innermost accretion flow around an astrophysical black hole. 
The~correct way to do this is {via} the induction equation, which in the case of an astrophysical accretion disk threaded by a large scale poloidal magnetic field ${\bf B}$ takes the form
\begin{equation}\label{induction}
\frac{\partial {\bf B}}{\partial t} = -\nabla\times \left( -{\bf v}\times {\bf B} + {\bf E}c +\eta\nabla\times {\bf B}\right)\ ,
\end{equation}
where  $\eta$ is the magnetic diffusivity in the interior of the disk. If $\nabla\times {\bf E}=0$, as is the case in stellar interiors, radiation does not generate electric currents neither magnetic fields. In stars, radiation pushes the electrons outwards leading to a surplus of electrons in the outer layers of the star, and a depletion in its center. As a result, stars become electrically polarized along their radius and develop an electric potential difference between the center and the surface (in the case of the sun, this is on the order of one volt), but no magnetic fields are generated. As we   now see, rotation introduces electric fields with non-zero {rotation (curl).} \footnote{Notice, however, that the contribution of rotation to the growth of the solar magnetic field by the Poynting--Robertson effect is insignificant \citep{CS66}.}

Combining the simple expression obtained in Equations~(\ref{PRforce}) and (\ref{E}), and the integrated form of Equation~(\ref{induction}), we obtain
\begin{equation}\label{inductionintegrated}
\frac{\partial \Psi}{\partial t} \approx 2\pi r\left( ({\bf v}\times {\bf B})^\phi - \frac{cf_{\rm rad}^\phi}{e} +\eta(\nabla\times {\bf B})^\phi\right)\ ,
\end{equation}
where  $\Psi$ is the magnetic flux contained inside radius $r$ (see also \citep{B-KLB02} for a generalized expression). The~second term in the {right-hand side} of the above equation generates a poloidal magnetic field in the direction of the angular velocity vector $\omega$ in the disk ($f_{\rm rad}^\phi$ is negative in most parts of the disk except possibly in its innermost part just above the horizon, as suggested by the results obtained in Figure~\ref{FigureSkyMaps}). The~magnetic flux thus generated closes in the outer parts of the disk not reached by radiation from the center where $f_{\rm rad}^\phi$ drops to zero. This poloidal magnetic flux will be advected inwards by the ideal accretion flow represented by the first term in Equation~(\ref{inductionintegrated}). We assume that, inside the ISCO, ideal MHD conditions apply, and flux accumulated inside the inner edge of the disk will keep growing. The growth will cease and the accumulated magnetic flux will saturate if the flow also carries along the return polarity of the magnetic field \citep{CK98, B-KLB02, CKC06}. However, if the return polarity lies in a region with significant magnetic diffusivity so that the third term in Equation~(\ref{inductionintegrated}) dominates over the first \citep{vB89, LPP94, LRN94} (see also, however,~\citep{LRB-K09}), the growth of the accumulated magnetic flux will proceed unimpeded. This latter point was emphasized by Contopoulos     and     Kazanas~(see Figure~1b in \citep{CK98}) and was missed by       {\citep{B-KLB02}}. As we show in the next section, actual accretion disks may favor both types of field evolution at different stages of their evolution (outward flux diffusion followed by the generation of poloidal magnetic field loops around the disk's inner edge, and inward flux advection followed by field reconnection with the flux accumulated inside the inner edge of the disk) .

The field growth cannot proceed beyond equipartition. There are various definitions of equipartition (e.g., balance of gravity by radiation pressure, balance of gravity by magnetic forces, etc.). Whatever the definition, it is clear that the field cannot keep growing steadily beyond its equipartition value $B_{\rm eq}$ at the inner edge of the disk, because the innermost accretion region will be severely disrupted, and our analysis will break down. Nevertheless, we can estimate a rough timescale $\tau_{\rm CB}$ for the innermost magnetic field to grow to astrophysically significant magnetic field values of order $B_{\rm eq}$ if we assume that the system radiates continuously at roughly its Eddington value. That~timescale may be obtained from a dimensional analysis of Equation~(\ref{inductionintegrated}) as
\begin{equation}\label{tauCB}
\tau_{\rm CB}\sim \frac{eB_{\rm eq}r_{\rm ISCO}}{cf_{\rm rad}^\phi}\ .
\end{equation}

The values of $\tau_{\rm CB}$ obtained for a $5M_\odot$ black hole with various black hole spin parameters and $B_{\rm eq}=10^7$~G  are shown in the fourth column of Table~\ref{table1}. These characteristic timescales vary from about one hour (for maximally rotating stellar mass black holes) to several days (for slowly rotating ones). These times scale roughly proportionally to $M^{3/2}$ with black hole mass, and therefore, the corresponding times to reach equipartition range from one to ten billion years for $10^8M_\odot$ supermassive black holes. We emphasize that these rough estimates have been obtained for accretion flows that radiate continuously at close to their Eddington limit. As we show in the next section, in realistic astrophysical sources such as XRB outbursts, this is not the case.

Observational confirmation of the Cosmic Battery in astrophysical systems (or any other kind of battery mechanism) may be found in observations of magnetic field asymmetries \citep{Cetal09, Ketal12, Cetal16, L13}. As is well known, the equations of motion in MHD involve only quadratic terms in the magnetic field through the Lorentz and electric forces $J\times B\sim (\nabla \times B)\times B\sim B^2$ and $\rho_e E\sim (\nabla\cdot E)E\sim E^2\sim B^2$, respectively. In other words, the dynamics of the flow do not depend on the direction of the magnetic field, unless some kind of battery mechanism is in action. {\citep{L13}} compared the direction of galactic rotation and the direction of the line-of-sight magnetic field in the central regions of nine spiral galaxies seen edge on. The~observations were found to be in agreement with the magnetic asymmetry {$B\parallel\omega$} predicted by the Cosmic Battery. Another asymmetry has to do with the helical structure of the magnetic field in the jet, and, in particular, with the direction of its axial electric current. The Cosmic Battery predicts that the axial electric current contained in the electron--proton disk wind/jet always flows away from the black hole (see Figure~2 of \citep{Cetal16} for details). A definite direction of axial electric current is related to a steady Faraday rotation measure gradient across the wind/jet. We were able to observe steady gradients across parts of the kpc-scale wind/jet in only 18 cases, and, in all of them, the direction of the axial electric current  was found to be outwards, in agreement with the prediction of the Cosmic Battery. For a more detailed review, the reader may consult \citep{C15}. 

\section{A New Paradigm}

We hope that we managed to convince the reader that the Cosmic Battery is one plausible origin for the large-scale magnetic field that threads the accretion--ejection flows around astrophysical black holes. Other options may be found in the recent work on mean-field dynamos in disks \citep{SFS14,Detal18,FG18}. In fact, the Cosmic Battery could be considered as the source for  the mean field dynamo in these simulations. 

We   now argue that the Cosmic Battery may be precisely the {\it missing key} that controls the general evolution of these systems. The ideal laboratory to test our ideas are X-ray binary outbursts where a multitude of temporal and spectral observations still awaits the development of a coherent self-consistent picture. Let us first consider the angular momentum conservation equation in the disk,~namely
\begin{equation}\label{angularmomentum}
\dot{M}_{\rm disk}\frac{\partial(rv_{\rm K})}{\partial r}+
\frac{\partial}{\partial r}(4\pi r^2 t_{r\phi}h)+
r^2 B_\phi B_z = 0\ .
\end{equation} 

Here, $t_{r\phi}$ is the tangential viscous stress   and $B_\phi, B_z$ are the toroidal and vertical components of the magnetic field at the base of the wind (notice that the product $B_\phi B_z$ is negative in a magnetic field configuration that removes angular momentum from the disk). Recent numerical simulations (see Figure~2 of {\citep{Setal16}}) suggest that the origin of the disk viscosity may be the Magneto-Rotational Instability (MRI) \citep{BH91}, and that %No mention of ref. 51, please revise.
\begin{equation}
t_{r\phi}\sim\frac{B^2}{4\pi}\ ,
\end{equation}
where $B$ is the average value of the magnetic field in the disk midplane. This numerical result allows us to simplify our analysis and to assume that the main factor that controls the removal of angular momentum in order for accretion to proceed is the large scale magnetic field that threads the accretion disk. In fact, Equation~(\ref{angularmomentum}) may be rewritten as
\begin{equation}\label{angularmomentum2}
\dot{M}_{\rm disk} \sim \frac{2r^2 B_\phi B_z}{v_{\rm K}}+\frac{2}{v_{\rm K}}\frac{\partial}{\partial r}(r^2 B^2 h)= \frac{\epsilon r^2 B^2}{v_{\rm K}}\ ,
\end{equation} 
where  $\epsilon$ is a factor of order unity.  {\citep{B19}} argued that $B^2$ (calculated in the interior of the disk) may be much larger than $|B_\phi B_z|$ (calculated on the surface of the disk). In the present paper, we   make the more natural assumption that the two are of the same order. In what follows, we   also drop factors of order unity in our calculations.

{In their seminal paper SS73,} Shakura and Sunyaev were the first to emphasize  the role of the magnetic field in the transport of angular momentum in astrophysical disks. They considered a turbulent magnetic field in the disk and incorporated its contribution to the disk viscosity in their now famous $\alpha$-parameter. Many years later, several researchers realized that a similar effect may be due to the action of a large scale magnetic field that threads the accretion disk, only now magnetic torques remove angular momentum to large distances (``infinity'') above and below the disk, and not radially through the disk. There exist many solutions in the literature in which the magnetic breaking mechanism is shown to work (e.g., \citep{FP95, L95}), but it seems that the astrophysical community still tries to ``do everything'' through the $\alpha$-parameter without specifically mentioning the magnetic field.

In a slight departure from the work of Ferreira and collaborators, we   consider the possibility that the distribution of magnetic field through the disk {\it may evolve} due to a slight imbalance between inward advection and outward diffusion through the disk, with one winning over large or small parts of the disk over the other. The timescale of this evolution is much longer than the dynamical timescale in the disk, and, therefore, the conclusions of previous works where the magnetic field was considered to be perfectly balanced in its position across the disk remain valid to a very good approximation.

Let us now discuss the evolution of a typical black-hole X-ray binary outburst, in particular GX 339-4 during its 2002--2003 outburst, through its associated Hardness-Intensity Diagram (HID). For this system,
\begin{equation}
M\approx 14\ M_\odot\ ,\ r_{\rm in}\approx 6GM/c^2=1.3\times 10^7\mbox{cm}\ ,\  \mbox{and}\ r_{\rm out}=1.5\times 10^9\ \mbox{cm}=100\ r_{\rm in}\ .
\end{equation}

\begin{figure}%[H]
\centering
\includegraphics[width=8 cm]{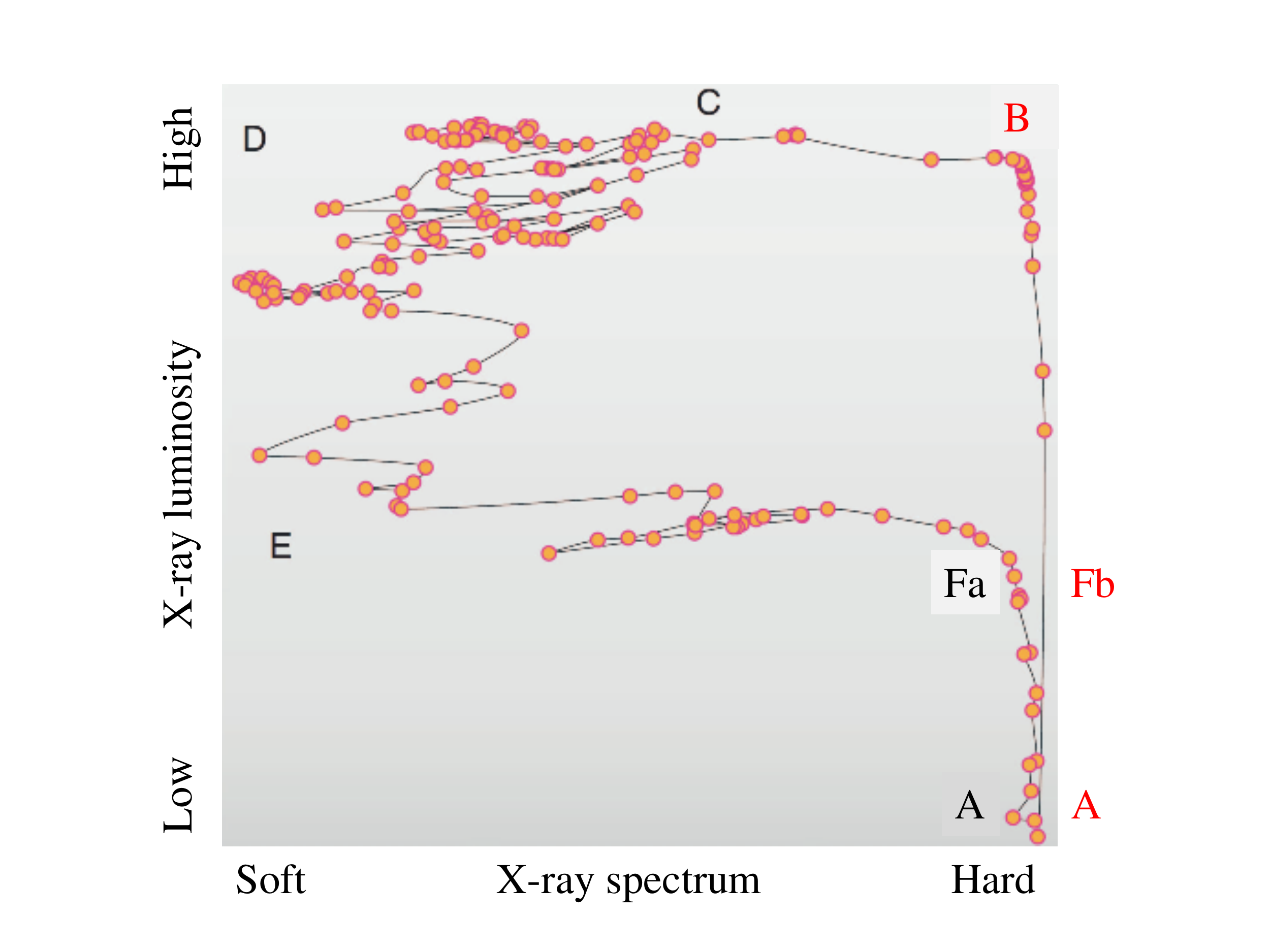}
\caption{Typical XRB Hardness-Intensity Diagram (HID) (shown the one for XRB GX 339-4 during its 2002--2003 outburst). {A: Quiescent state; B: High Hard state with fully developed compact jet; C: Bright Intermediate state where the compact jet is destroyed; D: Soft High state with no jet; E: Faint Intermediate state; Fa: Intermediate state traversed downwards (compact jet reappears but decreases in size); Fb: Intermediate state traversed upwards (compact jet grows in size). Red fonts: states where the Cosmic Battery operates} (adapted from \cite{FB12}).}
\label{FigureHID}%NOTE: Remove capitalization here and below?
\end{figure}

\begin{figure}%[H]
\centering
\includegraphics[width=11 cm]{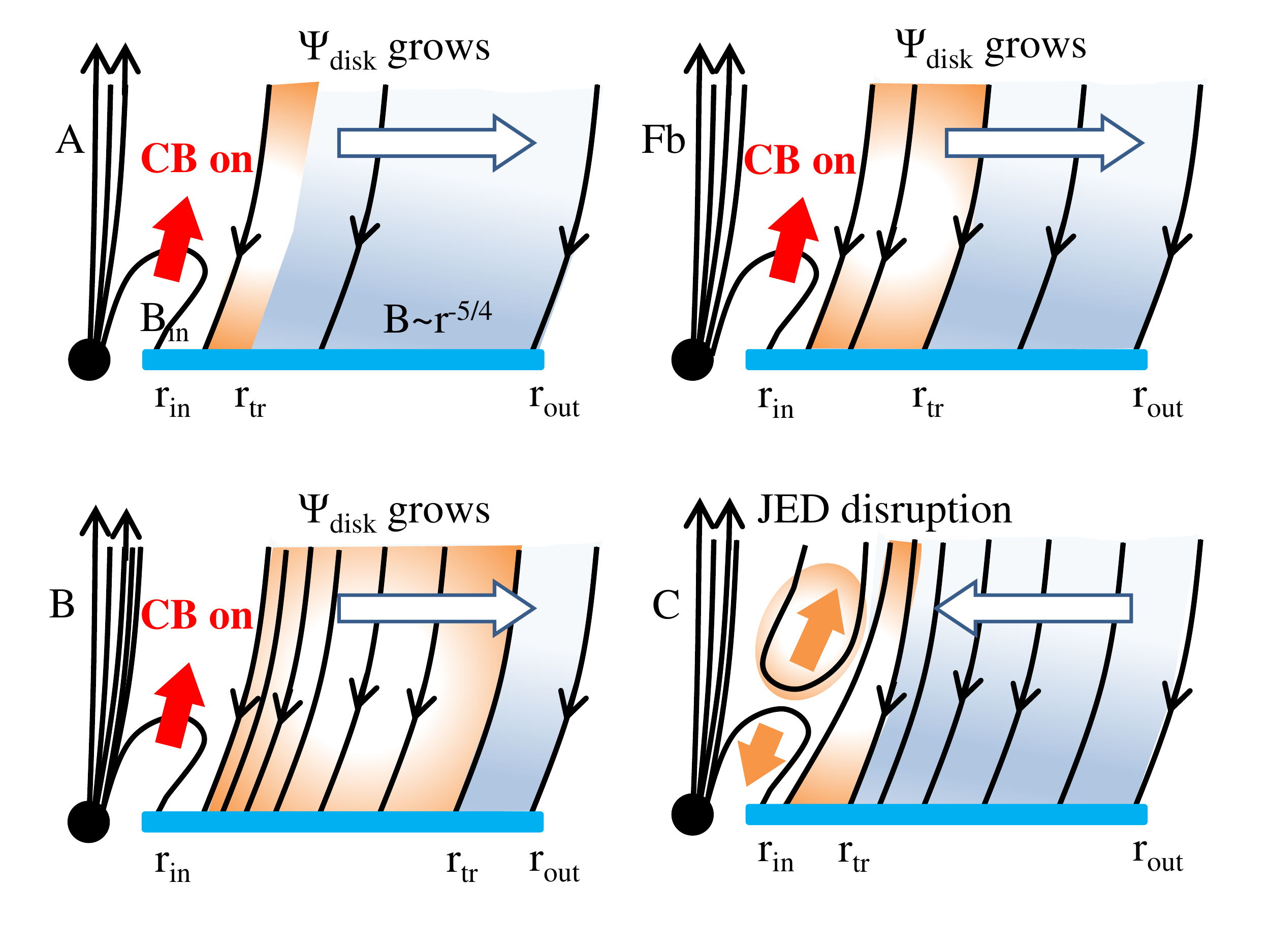}
\caption{{Schematic of our proposed model. Ascending phase of the HID. Blue strip: accretion disk; Black circle: central black hole; Solid lines with arrows: magnetic field. The direction of the angular velocity vector $\omega$ is along the $z$-axis. Magnetic field directions shown according to the predictions of the Cosmic Battery:
$B$ parallel to $\omega$ around the black hole, $B$ anti-parallel to $\omega$  in the disk. Blue arrows: opening up of poloidal loops generated by the Cosmic Battery; White arrows: direction of flux redistribution in the disk; Orange arrows: reconnecting magnetic flux;  Light blue: weak disk wind; Light orange regions: compact jet from JED; Light orange oval: ejected blobs of matter as the inner part of the JED is destroyed.}}
\label{FigureAscending}
\end{figure}

\begin{figure}[H]
\centering
\includegraphics[width=11 cm]{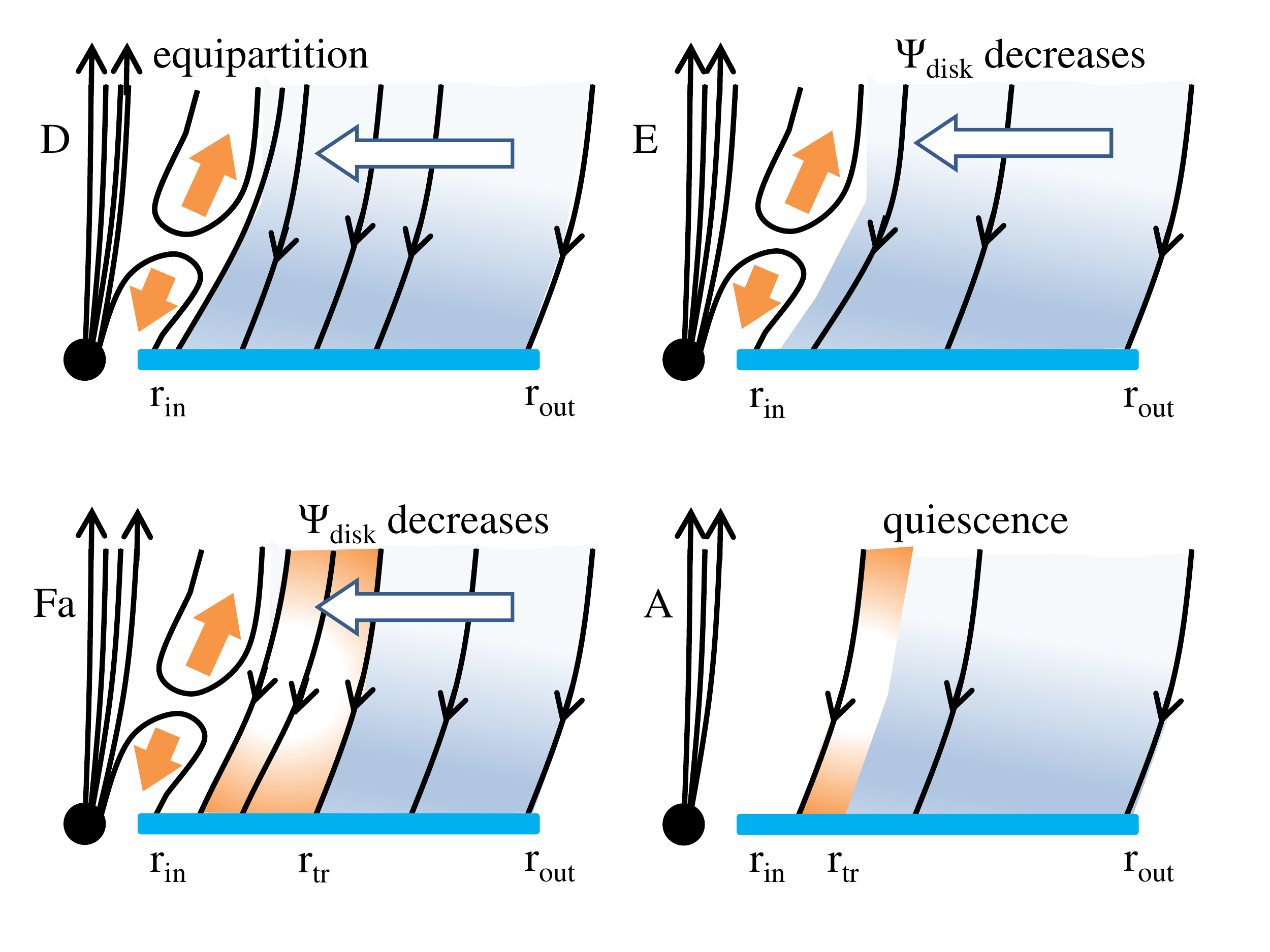}
\caption{Descending phase of the HID (similar to Figure~\ref{FigureAscending}). Notice how similar the intermediate states Fa and Fb are (they differ only in the direction of field redistribution through the disk). Notice also that the direction of flux redistribution during the ascending and descending phases is opposite to that in the model for XRB state transitions of \citep{BA14}.}
\label{FigureDescending}
\end{figure}

Five positions in the HID are very important during the system's evolution (see Figures~\ref{FigureHID}--\ref{FigureDescending}):
\newline
%\unskip

\noindent
A: The Quiescent state, which lasts for several months to a few years. In that state, the accretion disk is threaded by a large scale weak magnetic field that generates weak radio emission, probably also a weak magnetically driven wind, but no compact jet. In such a configuration, the mass loss in the wind is very weak, $\dot{M}_{\rm disk}\approx \mbox{const.}$, $\xi\approx 0$, thus Equation~(\ref{angularmomentum2}) yields \begin{equation}\label{BPscaling}
B(r)\approx B_{\rm in}\left(\frac{r}{r_{\rm in}}\right)^{-5/4}\ ,
\end{equation}
where  $B_{\rm in}\equiv B(r_{\rm in})$. This is the canonical radial magnetic field scaling first proposed in BP82. In that case, the total magnetic flux threading the accretion disk that extends from $r=r_{\rm in}$ to $r_{\rm out}$ is equal to
\begin{equation}\label{Psidisk}
\Psi_{\rm disk}=\int_{r_{\rm in}}^{r_{\rm out}}2\pi r B(r){\rm d}r\approx
\frac{8\pi}{3}B_{\rm in} r_{\rm in}^{2}\left(\frac{r_{\rm out}}{r_{\rm in}}\right)^{3/4}\approx 260\ B_{\rm in} r_{\rm in}^{2}
\end{equation}

In the present work, we   assume that $\xi$ remains always very close to zero, and that the radial field configuration always remains close to the canonical BP82 one. Notice that the investigation of BP82-type Magnetized Accretion--Ejection Structures (MAES) by Ferreira and collaborators required that the magnetic field stays in place and does not diffuse inwards or outwards through the accreting flow. In reality, the total magnetic flux threading the disk may slowly change as the system evolves in the HID, and, therefore,     to keep the BP82 radial scaling, the field will re-arrange itself as the system evolves during the XRB outburst. We conclude that, in practice, the delicate balance between inward advection and outward diffusion is not   one hundred percent satisfied. We   show below why this is very important during an XRB outburst, and how it {may be} related to the action of the Cosmic Battery around the central black hole.\newline

\noindent
B: The High Hard state. At some point in time, without any prior indication, the system ``decides'' to flare up. It takes a few months for the outburst to rise to its highest luminosity. During that time, a compact jet appears, whose radius at its base seems to increase with time as the system evolves from the quiescent to the high-hard state \citep{KR18}. Quoting {\citep{Fetal06}}, ``whenever a disk is capable of driving jets, these will carry away a fraction of the released accretion (gravitational) energy, and the disk luminosity will be quenched (it radiates only a small fraction of the accretion power)''. It is well known that we can interpret the observed spectra and spectra variations during an XRB outburst with a low radiative efficiency ADAF disk model inside an evolving transition radius $r_{\rm tr}$ (e.g., \citep{NMcCY96, Hetal97, Eetal98, Eetal01}). Ferreira and collaborators argued that it is possible to interpret the observations equally well with their JED model \citep{Fetal06, Metal18a, Metal18b}. We prefer the latter explanation since, as we show below, it requires a minimum number of assumptions and free parameters. Notice that the rise from the Quiescent to the High state must be~explained.\newline

\noindent
C: The Bright Intermediate state. The system transitions from the Hard to the Soft High state, and the radius of the compact jet quickly diminishes in size at its base \citep{KR18}. We associate this with a reduction of the size of the inner JED. At some point, the jet disappears abruptly and episodically in the form of a micro-quasar. According to the theory of JEDs, this corresponds to an abrupt reduction of the magnetization in the jet. Obviously, if the system is threaded by a large scale unidirectional magnetic field, there is no way to make the magnetic flux disappear. On the contrary, if magnetic flux is carried inwards by the shrinking JED, the magnetic field and its associated magnetization are both expected to increase and not decrease. This sudden reduction of the magnetization in the inner accretion disk must be explained.\newline

\noindent
D: The Soft High state. In that state, the JED has disappeared, and the soft emission originates in a disk of Shakura--Sunyaev type. The system stays in the soft state but gradually decreases in luminosity in the course of several months. This decrease in luminosity is not monotonic, and the system makes several unsuccessful efforts to return to the hard state. This non-monotonic decrease in luminosity must be explained.\newline

\noindent
E: The Faint Intermediate state. After several months, the system transitions to a state beyond which the system is ``ready'' to return to the Hard state without hesitating. It is interesting that a particular bursting XRB system may reach differing levels of peak luminosity in different outbursts, but always returns to approximately the same Faint Intermediate state before it decides to return to the quiescent state in the HID. This observation must be explained.\newline

\noindent
F: The Low Hard state. The return point. The system reaches the Low Hard state and then decides to turn downwards in the HID and return to the quiescent state. Position Fa during the descending phase is very close to position Fb during the ascending crossing of the Low Hard state. The system configuration must be very similar at Points Fa and Fb (similar spectra and luminosities, similar disk types, similar radio emissions, and similar distributions of large-scale magnetic field through the disk), yet in the former the system traverses the HID downwards, whereas in the latter it traverses it upwards. Why that happens remains   unexplained.\newline

We believe that the answer to the latter question holds the key to understanding the dynamics of Magnetized Accretion--Ejection Structures. We propose that the answer has to do with the delicate balance (or better {\it imbalance}) between inward flux advection and outward flux diffusion. In our physical picture, the action (or inaction) of the central Cosmic Battery plays a key role as follows:\newline

The Cosmic Battery generates the magnetic field in the immediate vicinity of the central black hole. One polarity of the field is held by the accreting matter infalling onto the central black hole, whereas the return polarity diffuses outward through the outer accretion disk. Let us denote the rate of return field generation by the Cosmic Battery as $\dot{\Psi}_{\rm CB}$. Obviously, $\dot{\Psi}_{\rm CB}$ depends on the innermost disk luminosity $L$, which depends on the accretion rate in the disk (see below). The pressing question is to understand why the system beyond Fa  evolves in the ascending part, whereas beyond Fb in the descending. We propose that the answer lies with the direction of magnetic flux redistribution: in the ascending part of the HID, outward diffusion slightly wins over inward advection, and, therefore, the disk is slowly filled with magnetic flux generated by the CB. According to the discussion in the previous section, the CB yields an approximate rate of generation of magnetic flux in the disk
\begin{equation}\label{PsidotCB}
\dot{\Psi}_{\rm CB}\approx \frac{L\sigma_T}{4\pi r_{\rm in}^2 ec}\left(\frac{v_{\rm K}(r_{\rm in})}{c}\right)2\pi r_{\rm in}c=f_{\rm CB}\frac{\dot{M}_{\rm disk}c^2\sigma_T}{2r_{\rm in}e}\ ,
\end{equation}
where the factor in front is defined as $f_{\rm CB}\equiv (L/\dot{M}c^2)(v_{\rm K}(r_{\rm in})/c)$. Combining Equations~(\ref{angularmomentum2}), (\ref{Psidisk}) and~(\ref{PsidotCB}) then yields
\begin{equation}
\frac{\partial B_{\rm in}}{\partial t}\approx \frac{f_{\rm CB}c^2 \sigma_T}{8e r_{\rm in}^{1/4}r_{\rm out}^{3/4} v_{\rm K}(r_{\rm in})}B_{\rm in}^2\ .
\end{equation}

This equation has the solution
\begin{equation}\label{Binasc}
B_{\rm in}(t)=\frac{B_{\rm quiesc}}{1-t/\tau_{\rm asc}}\ ,
\end{equation}
where the time $t$ is measured from the beginning of the outburst in the quiescent state, and the characteristic ascending timescale $\tau_{\rm asc}$ is equal to
\begin{equation}
\tau_{\rm asc} \approx \frac{8 e r_{\rm in} v_{\rm K}(r_{\rm in})}{f_{\rm CB}c^2 \sigma_T B_{\rm quiesc}}\left(\frac{r_{\rm out}}{r_{\rm in}}\right)^{3/4}=\frac{1}{f_{\rm CB}}\left(\frac{B_{\rm quiesc}}{10^6\ \mbox{G}}\right)^{-1}\ \mbox{yr}\ .
\end{equation}

This timescale is much longer than the characteristic timescales of Table~\ref{table1} because according to our present model, most of the time, the system accretes at rates two orders of magnitude below equipartition, thus most of the time, the Cosmic Battery operates very inefficiently. Notice that the imbalance between inward flux advection and outward flux diffusion is just a very small fraction of each one of these terms, namely
\begin{equation}
\frac{\dot{\Psi}_{\rm CB}}{2\pi r_{\rm in}v_r(r_{\rm in})B_{\rm in}}=f_{\rm CB}\frac{B_{\rm in}c^2 \sigma_T}{2\pi ev_{\rm K}(r_{\rm in})v_r(r_{\rm in})}\lsim
 10^{-8} \left(\frac{B_{\rm in}}{10^7\ \mbox{G}}\right)\ .
\end{equation}

It is impossible to simulate such a small effect in global numerical simulations like the ones performed by       {\citep{TNMcK11}}. In the ascending phase of the HID (the growing phase of the outburst), the disk is slowly filled with magnetic flux, and the field at its inner edge becomes stronger and stronger with time according to Equation~(\ref{Binasc}). As a result, the disk accretion rate and luminosity also rise as
\begin{equation}\label{Mdotasc}
L(t)\propto \dot{M}_{\rm disk}(t)\sim \frac{r_{\rm in}^2 B_{\rm quiesc}^2}{v_{\rm K}(r_{\rm in})(1-t/\tau_{\rm asc})^2}\ .
\end{equation}

At the same time, the magnetic flux $\Psi_{\rm BH}$ accumulated around the central black hole continuously~increases.
%In the Appendix, we confirm that $\Psi_{\rm BH}$ can be held by the accretion flow, i.e. that the outward force it excerts on the inner accretion flow does not exceed gravity. 
The accretion rate becomes stronger and stronger according to Equation~(\ref{angularmomentum2}). At some point, the accretion will approach equipartition with radiation and the innermost accretion flow will be disrupted. Equipartition corresponds to
\begin{equation}
B_{\rm eq}\approx 
\left(\frac{4\pi GM m_{\rm p}v^2_{\rm K}(r_{\rm in})}{f_{\rm CB} r_{\rm in}^2 \sigma_T c^2}
\right)^{1/2}\approx 7\times 10^6\ f_{\rm CB}^{-1/2}\ \mbox{G}\ .
\end{equation}

Notice that, according to Equation~(\ref{Mdotasc}), as the luminosity rises by two orders of magnitude with respect to the quiescent state, the magnetic field rises only by one order of magnitude. Therefore, we naturally expect that the innermost value of the magnetic field in the disk during quiescence is on the order of
\begin{equation}
B_{\rm quiesc}\approx \frac{B_{\rm eq}}{10}\sim 10^6\ \mbox{G}\ .
\end{equation}

Beyond that point, the inner accretion flow is disrupted, the Cosmic Battery ceases to operate, and no new magnetic  flux is generated. The delicate balance between inward flux advection and outward flux diffusion is now reversed, and the magnetic flux in the disk is gradually brought to the center where it reconnects with the magnetic flux accumulated around the central black hole. As a result, in the descending return part of the HID, inward advection slightly wins over outward diffusion, and the total flux in the disk decreases. 
%at a rate roughly dictated by the reconnection rate at the inner part of the disk. Using Equation~(\ref{angularmomentum2}), this is roughly given by
%\begin{equation}\label{Psidotrec}
%\dot{\Psi}_{\rm rec}\approx -2\pi r_{\rm in}v_{\rm A}B_{\rm in}\sim -2\pi fr_{\rm in}v_{\rm K}(r_{\rm in})B_{\rm in}
%\end{equation}
%where the factor $f$ is equal to $f\sim (h/r)^{1/2}(v_r/v_{\rm K})^{1/2}$. Equation~(\ref{Psidisk}) then yields
%\begin{equation}
%\frac{\partial B_{\rm in}}{\partial t}\sim -\frac{fv_{\rm K}(r_{\rm in})}{\pi r_{\rm in}^{1/4}r_{\rm out}^{3/4}}B_{\rm in}\ .
%\end{equation}
%This has the solution
%\begin{equation}
%B_{\rm in}=B_{\rm eq}\exp\left(-\frac{t}{\tau_{\rm desc}}\right)\ .
%\end{equation}
%Here, the time $t$ is measured from the beginning of the decreasing phase of the outburst, and the characteristic descending timescale $\tau_{\rm desc}$ is roughly equal to
%\begin{equation}
%\tau_{\rm desc} \sim \frac{\pi r_{\rm in}}{fv_{\rm K}(r_{\rm in})}\left(\frac{r_{\rm out}}{r_{\rm in}}\right)^{3/4}\approx\frac{0.1}{f}\ \mbox{s}\ .
%\end{equation}
%In the descending phase of the HID, the overall field in the disk becomes weaker and weaker, because magnetic flux is continuously anihilated around the central black hole. As a result, the disk accretion rate and luminosity also decrease as
%\begin{equation}\label{Mdotdesc}
%L(t)\propto \dot{M}_{\rm disk}(t)\sim \frac{r_{\rm in}^2 B_{\rm eq}^2}{v_{\rm K}(r_{\rm in})}\exp\left(-\frac{2t}{\tau_{\rm desc}}\right)\ .
%\end{equation}
The overall field in the disk becomes weaker and weaker, and, as a result, the disk accretion rate and luminosity also decrease. 
During the descending phase, the system remains close to equipartition. This keeps disrupting the innermost accretion flow. At the same time, the field annihilation proceeds through a series of field reconnection events, and is therefore, not clearly monotonic. This explains why the decrease of the outburst luminosity is rather irregular and often non-monotonic.

As in the ascending phase, the magnetic field in the descending phase is strong enough to be able to support a JED at the inner part of the disk.  However, because the innermost accretion flow is disturbed so dramatically around equipartition, the conditions for restarting the inner compact jet are not favorable. Eventually,  the central magnetic field decreases well below equipartition so that the JED forms without further disruption. Why this takes place at that particular value of the disk luminosity that corresponds to Point E in the HID is not yet clear to us. We need to investigate in greater detail what happens to the overall accretion--ejection around equipartition, possibly with numerical simulations.

After the compact jet reappears, the disk magnetic flux continues to ``drip'' towards the inner edge of the disk, and, therefore, the Cosmic Battery does not operate to halt the field decay and reverse the decrease of the disk accretion rate. This is why at Point Fa  the system continues to traverse the HID downwards towards the quiescent state, and does not start a new outburst as is the case at Point Fb where the system traverses the HID upwards.


%%%%%%%%%%%%%%%%%%%%%%%%%%%%%%%%%%%%%%%%%%
\section{Summary and Conclusions}

We believe that the key element missing from most previous efforts to understand the dynamics of accretion--ejection flows in X-ray binary outbursts may be the action of the Cosmic Battery in the innermost accretion region around the central black hole. Our model is plausible, but has not yet been conclusively proven by simulations (see however \citep{Cetal18}). The Cosmic Battery offers the possibility to generate poloidal magnetic field loops around the inner edge of the accretion disk. Due to differential rotation, these loops open up above and below the disk. As a result, one polarity of the field (the one where $B$ and $\omega$ are parallel) is brought to the center and inundates the black hole horizon, whereas the return field polarity (the one where $B$ and $\omega$ are anti-parallel) threads the surrounding disk. 

What is crucial for the distribution of magnetic flux in the disk is the delicate balance between inward flux advection and outward flux diffusion. We propose that the field reaches such a balance in the disk and therefore attains a large scale configuration similar to the one proposed by {\citep{BP82}}. The~balance, however, is not perfect and is influenced by what happens around the center, namely by the action of the Cosmic Battery. 

During the ascending part of the XRB outburst, the Cosmic Battery is in operation and magnetic flux is introduced to the accretion disk at its inner edge. As a result, the total magnetic flux that threads the disk continuously increases at a particular rate given by Equation~(\ref{PsidotCB}). Similarly, the disk accretion rate and its associated luminosity also increase according to Equation~(\ref{Mdotasc}). Eventually, the accretion rate reaches equipartition with radiation, at which point the innermost accretion flow is dramatically disturbed. The outward field diffusion is reversed, and magnetic flux is advected inwards towards the central black hole. The Cosmic Battery ceases to operate, and the total magnetic flux that threads the accretion disk is gradually lost via reconnection with the magnetic field that is accumulated around the center. At the same time, the energetic jet is destroyed and the accretion disk transitions to a Shakura--Sunyaev type. The total magnetic flux that threads the disk continuously decreases, as does the disk accretion rate and its associated luminosity. Notice that the directions of flux redistribution during the ascending and descending phases of the outburst that we propose in the present work are {\it opposite} to those proposed by       {\citep{BA14}} in their model for XRB state transitions. In both models, however, the accumulated magnetic field increases/decreases during the ascending/descending phase of the outburst. The difference between the two models is precisely the operation of the {Cosmic B}attery around the central black hole.%NOTE: Please confirm where such capitalization is necessary, use uniformly and remove where appropriate.

It is interesting that, during the descending phase of the outburst, the system may make several attempts to generate a compact jet, i.e. to form a JED in its innermost part. This may happen several times as the luminosity decreases, and a transient compact jet may form several times before the system eventually reaches the so-called {Faint I}ntermediate state (point E in the HID) where it makes a final transition to a JED and subsequently returns to the quiescent state. According to {\citep{Fetal06}},   for a JED to form, the magnetization $\mu$ in the disk, defined as the ratio of magnetic to gas pressures must be greater than unity. In our present formulation, and with the help of Equation~(\ref{angularmomentum2}), 
\begin{equation}
\mu\equiv \frac{B^2/8\pi}{\rho_{\rm disk} c_s^2}\sim
\frac{\dot{M}_{\rm disk}v_{\rm K}/(\epsilon r^2)}{(\dot{M}_{\rm disk}/2\pi r 2h v_r) (h v_{\rm K}/r)^2}\approx
\frac{v_r}{v_{\rm K}}\cdot\frac{r}{h}
\end{equation}

The magnetization parameter $\mu$ may exceed unity (within detailed numerical factors that are missing from our crude calculation), either when the accretion velocity approaches the Keplerian rotational velocity and $h\sim r$, either when $h\ll r$. The latter is a possibility in a Shakura--Sunyaev-type disk. We are more interested here in the former possibility where $v_r\sim v_{\rm K}$. It seems to us that the system can always support an inner JED, but when it reaches equipartition, the JED is destroyed. However, whenever it finds the opportunity, as in the multiple  attempts of the system to cross the so-called jet line in the HID, the field is strong enough for a JED to form. Why the disk changes type and generates a persistent (not transient) JED in its innermost parts at the particular luminosity that corresponds to point E in the HID is not {yet} clear to us (see, however, also \citep{BA14, K15} for interesting ideas).

We acknowledge that much more work must be done   to understand all the details of the dynamics of bursting XRB systems. We have     not investigated ourselves how the various spectral states of the HID arise in our picture of an inner JED and an outer Shakura--Sunyaev-type disk, and only refer to the work of other groups. The assumption that accretion proceeds only via magnetic torques due to the large scale magnetic field that threads the disk is clearly simplistic. This allows us, however, to make very definite simple predictions about the rise and fall of the outburst that can be directly compared to observations. Finally, our treatment of the Cosmic Battery is too   simplistic since it ignores general relativity and the complicated geometry of the central source of radiation, and assumes a point central source of radiation as in the Poynting--Robertson effect in our solar system.

We would like to note that our results are also applicable to AGN with or without jets. It is possible that, as in the case of bursting XRB sources,  jets appear only during a small fraction of the system's lifetime. This may be the answer why only a small fraction of AGN develop large scale jets. The scales of those systems are much larger, namely $M\sim 10^{9}\ M_{\odot}$, $r_{\rm in}\sim 10^{15}\ \mbox{cm}$, and therefore, $B_{\rm eq}\sim 10^3\ \mbox{G}$. If $B_{\rm quiesc}\sim 10^3\ \mbox{G}$, then $\tau_{\rm asc}\sim 10^{11}\ \mbox{yr}$, which is longer than the Hubble time. Obviously, the model parameters that we used in GX 339-4 (where most of the time the system accretes several orders of magnitude below equipartition and the Cosmic Battery operates very inefficiently) do not directly apply here. We expect that AGN with jets accrete at rates close to equipartition. More work needs to be done to understand the role of the Cosmic Battery in AGN with jets.

In summary, we hope that we have convinced the reader that the action (or inaction) of the Cosmic Battery in the central regions of XRB and AGN may be {\it the missing key} in the study of the dynamics of outbursts in these systems.

%\begin{listing}[H]
%\caption{Title of the listing}
%\rule{\textwidth}{1pt}
%\raggedright Text of the listing. In font size footnotesize, small, or normalsize. Preferred format: left aligned and single spaced. Preferred border format: top border line and bottom border line.
%\rule{\textwidth}{1pt}
%\end{listing}
\vspace{6pt}

\funding{This research received no external funding.}

%%%%%%%%%%%%%%%%%%%%%%%%%%%%%%%%%%%%%%%%%%
\acknowledgments{Most of the ideas presented in this work arose from discussions with Nick Kylafis, Jonathan Ferreira, Demos Kazanas, and Sergei Bogovalov. The material in Section~{2} was provided by Leela Koutsantoniou.}%NOTE: Section not in \ref format.

%%%%%%%%%%%%%%%%%%%%%%%%%%%%%%%%%%%%%%%%%%
\conflictsofinterest{The author declares no conflict of interest.}

%=====================================
% References, variant A: internal bibliography
%=====================================
\reftitle{References}

% The following MDPI journals use author-date citation: Arts, Econometrics, Economies, Genealogy, Humanities, IJFS, JRFM, Laws, Religions, Risks, Social Sciences. For those journals, please follow the formatting guidelines on http://www.mdpi.com/authors/references
% To cite two works by the same author: \citeauthor{ref-journal-1a} (\citeyear{ref-journal-1a}, \citeyear{ref-journal-1b}). This produces: Whittaker (1967, 1975)
% To cite two works by the same author with specific pages: \citeauthor{ref-journal-3a} (\citeyear{ref-journal-3a}, p. 328; \citeyear{ref-journal-3b}, p.475). This produces: Wong (1999, p. 328; 2000, p. 475)

%=====================================
% References, variant B: external bibliography
%=====================================
%\externalbibliography{yes}
%\bibliography{your_external_BibTeX_file}

\end{document}